\documentclass{llncs}
\usepackage{xcolor}

\usepackage{graphicx}
\usepackage{amsmath,amssymb}
\usepackage{hyperref}
\usepackage{placeins}

\title{A Study on Malicious Browser Extensions in 2025}

 \author{
 Shreya Singh \and
 Gaurav Varshney \and
 Tarun Kumar Singh \and
 Vidhi Mishra \and
 Khushi Verma
 }

 \institute{
 Department of CSE, IIT Jammu, Jammu, India \\
 \email{
 \{2022pct0019, gaurav.varshney, 2022pct0020, 2023pcs0043, 2022ucs0094\}@iitjammu.ac.in
 }
 }

\begin{document}
\maketitle

\begin{abstract}
Browser extensions are additional tools developed by third parties that integrate with web browsers to extend their functionality beyond standard capabilities. However, the browser extension platform is increasingly being exploited by hackers to launch sophisticated cyber threats. These threats encompass a wide range of malicious activities, including but not limited to phishing, spying, Distributed Denial of Service (DDoS) attacks, email spamming, affiliate fraud, malvertising, and payment fraud. This paper examines the evolving threat landscape of malicious browser extensions in 2025, focusing on Mozilla Firefox and Chrome. Our research successfully bypassed security mechanisms of Firefox and Chrome, demonstrating that malicious extensions can still be developed, published, and executed within the Mozilla Add-ons Store and Chrome Web Store. These findings highlight the persisting weaknesses in browser's vetting process and security framework\cite{BrowserMarketShareReference}. It provides insights into the risks associated with browser extensions, helping users understand these threats while aiding the industry in developing controls and countermeasures to defend against such attacks. All experiments discussed in this paper were conducted in a controlled laboratory environment by the researchers, adhering to proper ethical guidelines. The sole purpose of these experiments is to raise security awareness among the industry, research community, and the general public.
\end{abstract}


\section{Introduction}
The rapid evolution of web browsers has transformed them from simple tools for accessing websites into sophisticated platforms that support a wide range of functionalities. Modern browsers such as Google Chrome, Mozilla Firefox, and Microsoft Edge not only enable seamless web browsing but also offer features like bookmarks, history tracking, and built-in security enhancements. However, users often demand additional capabilities that extend beyond these native features, driving the widespread adoption of browser extensions. These extensions, developed by third parties, allow users to customize and enhance their browsing experiences, providing tools such as ad blockers, password managers, and productivity boosters.

Despite their utility, browser extensions represent a double-edged sword. While they significantly enhance user convenience, they also introduce numerous security vulnerabilities. Malicious actors exploit browser extensions to execute sophisticated attacks, including phishing, keylogging, spying, data theft, and session hijacking~\cite{ref10}. This dual nature—offering benefits while posing significant risks—has made browser extensions a critical focus of cybersecurity research.

This paper aims to study and experimentally validate the inherent malicious capabilities of browser extensions, with a particular focus on  Mozilla Firefox and Chrome, the world’s most widely used browser \cite{BrowserMarketShareReference}. By analyzing real-world examples and examining attack techniques employed by malicious extensions, the study highlights the evolving threat landscape posed by browser extensions in the year 2025. The experiments and findings presented in this paper are designed to assist internet researchers and users in understanding these threats, while also guiding the industry toward the development of effective controls and countermeasures to mitigate risks associated with browser extensions.

Ultimately, this research underscores the critical need for secure practices and policies to protect against browser extension attacks. It also emphasizes the importance of striking a balance between providing APIs for extended functionality and addressing the security risks they introduce, which can often go undetected. While this study focuses on Google Chrome and  Mozilla Firefox, most modern browsers, including Microsoft Edge, Safari, support extensions built on standardized APIs. Many of the attack techniques explored in this paper could be adapted to other browsers, suggesting a broader industry-wide risk.

\subsection{Browser Extensions}
Browser extensions \cite{ref2},\cite{ref3}, developed by third-party creators, enhance browser functionality by adding diverse features. However, they also present significant security risks, as malicious actors exploit them to carry out sophisticated attacks, including phishing, spying, DDoS, email spamming, affiliate fraud, malvertising, and payment fraud. This paper investigates the vulnerabilities of current browsers, with a focus on  Mozilla Firefox and Google Chrome due to its widespread use and popularity. It explores the technical weaknesses in the extension platform that attackers leverage to execute such malicious activities. By analyzing these vulnerabilities, the study highlights potential improvements, such as enhanced vetting processes for extensions, increased user awareness, and stronger security protocols within browsers. Implementing these measures can mitigate risks, enhance browser security, and ensure a safer browsing experience for users.

\subsection{Malicious Browser Extension Based Attacks }
MBEs are third-party add-ons that exploit browser APIs to perform unauthorized actions such as data theft, spying, content manipulation, and session hijacking. Increasingly, these extensions, often disguised as productivity tools, enable attacks including \textit{keylogging}, \textit{data exfiltration}, \textit{malicious code injection}, and \textit{session hijacking}. Early MBEs (2013–2015) primarily focused on basic data theft—\textit{keylogging} (e.g., \textit{HoverZoom, 2013}) and activity tracking (e.g., \textit{BBC News Reader, 2014})—eroding user trust. By 2016, attackers employed more advanced techniques such as \textit{web request interception} (e.g., \textit{iCalc, 2016}) and \textit{token theft via phishing} (e.g., \textit{Viralands, 2016}). From 2017 onward, MBEs adopted complex payloads for financial and operational exploitation, including \textit{ad fraud} (e.g., \textit{Web Developer, 2017}) and \textit{proxy-based traffic manipulation} (e.g., \textit{Dubbed Copyfish, 2017}). Recently, supply chain attacks (e.g., \textit{Cyberhaven, 2024}) have emerged, where adversaries compromise trusted developer accounts using social engineering and OAuth token abuse to deploy malicious updates. Table~\ref{tab:browser-attacks} summarizes notable MBE incidents from 2013 to 2024, illustrating the shift from opportunistic attacks to coordinated, high-value campaigns exploiting browser vulnerabilities.

\begin{table*}[h]
\caption{Malicious Extension Based Attacks (2013-2024) 
}
\label{tab:browser-attacks}
\centering
\renewcommand{\arraystretch}{1} 
\resizebox{\textwidth}{!}{%
\small 
\begin{tabular}{|l|l|p{9cm}|}
\hline
\textbf{Extension (Year)} & \textbf{Attack Type} & \textbf{Description} \\ \hline
HoverZoom (2013) \cite{ref9} & Keylogging & Browse images on websites by hovering. Collecting online form data and selling users’ keystrokes. \\ \hline
Tweet This Page (2014) & Content Injection & Tweet a Page. Turned into an ad-injecting machine; started hijacking Google searches. \\ \hline
BBC News Reader (2014) & Spying & Get latest news and articles. Tracks user browsing data. \\ \hline
Autocopy (2014) \cite{ref9} & Spying & Select text and automatically copy to the clipboard. Sends a lot of user data back to its servers. \\ \hline
Hola Unblocker (2015) & DDoS & Easy-access to region blocked content. Bandwidth from users being sold to cover costs (powers botnets for attack). \\ \hline
Marauder’s Map (2015) \cite{ref9} & Spying & Plot your friends’ location data from Facebook on a map. A hacker can know if you’re not home, shops you visit frequently, who you spend most time with. \\ \hline
Viralands (2016) \cite{ref9} & Phishing & “Verify your age” to access restricted content. Access to Facebook access token; login credentials stolen. \\ \hline
iCalc (2016) \cite{ref9} & Webpage Manipulation & Functional Calculator. Creates a proxy and intercept web requests, taking commands and updates from a domain. \\ \hline
Dubbed Copyfish (2017) \cite{ref9} & Mal-Ads & Extract text from images, PDFs, videos. Equipped with ad injection capabilities. \\ \hline
Web Developer (2017) & Affiliate Fraud & Adds a toolbar button to the browser with web developer tools. Substitute ads on browser, hijacking traffic from legit ad networks. \\ \hline
Nano Adblocker/Nano Defender (2020) \cite{soyacincau2020} \cite{arstechnica2020} \cite{ref9}& Spying & Adblocker. Collected user data and sent it to remote servers. \\ \hline
The Great Suspender (2021) \cite{thehackernews2021} \cite{ref9} & Malware & Suspends unused tabs to save memory. Injected malicious code to steal data. \\ \hline
SessionManager (2022)  & Data Theft & Manage browser sessions. Stole session cookies and other data. \\ \hline
Sakula Rat (2023)  & Remote Access Trojan & Used for APT campaigns. Allowed remote control and data exfiltration from infected browsers. \\ \hline
Session Stealer (2023) & Hijacking & Manage browser sessions. Stole active session cookies to hijack accounts. \\ \hline
Cyberhaven (2024) \cite{thehackernews2024} \cite{bleepingcomputer2024chromeextensions} \cite{fieldeffect2025} & Supply Chain & Compromised via phishing targeting developer accounts. Distributed malicious versions, stealing Facebook access tokens and bypassing 2FA. \\ \hline
StealthSpy (2024) \cite{thehackernews2024}  & Spying & Disguised as a productivity enhancer. Secretly records browsing history and keylogs sensitive data. \\ \hline
AdSkimmer Pro (2024) \cite{thehackernews2024}  & Ad Fraud & Claimed to block ads. Injected its own ads and skimmed affiliate commissions from legitimate sites. \\ \hline
QuickAccess Helper (2024) \cite{thehackernews2024} & Phishing & Promised faster access to commonly visited sites. Redirected users to phishing pages that stole credentials. \\ \hline

\end{tabular}%
}

\end{table*}

\section{Literature Review on MBEs}
Malicious browser extensions (MBEs) increasingly threaten user privacy by masquerading as benign productivity tools while covertly stealing data, injecting ads, or hijacking sessions. Studies such as ParaSiteSnatcher \cite{ParaSiteSnatcher} and Cyberhaven’s Chrome extension \cite{cyberhaven2025,thehackernews2024} illustrate the multifaceted impact of these threats. Varshney et al. \cite{ref9} demonstrated that the inherent access to sensitive browser APIs enables attackers to execute phishing, spying, DDoS, email spamming, and affiliate fraud. This vulnerability is further examined in \cite{cyberattack}, which highlights how Chrome’s open extension platform facilitates cyberfraud and cyberspying. Chang and Chen’s work \cite{ref10} emphasizes the risks of runtime information leakage from extensions, while Maunder \cite{maunder2017extensionattacks} reveals that millions of malicious extensions can operate undetected. At DEF CON 32, SquareX \cite{squarex2024} showed that even with Google’s Manifest V3 framework, MBEs can bypass security measures, stealing live streams, cookies, and user credentials. Collectively, these studies underscore the need for enhanced detection strategies and stricter controls, as the evolving threat landscape suggests that MBEs will continue to pose significant risks into 2025.

 Several studies have proposed techniques to detect and mitigate malicious browser extensions. Wang et al. \cite{wang2018detect} introduced a machine-learning model that combines static and dynamic analyses of JavaScript, HTML, and CSS to classify extensions with over 95\% accuracy. Kaushik et al. \cite{kaushik2021safeguarding} advocate for continuous monitoring through enhanced permission management, API tracking, and real-time behavior analysis to preemptively block harmful extensions. Varshney and Misra \cite{varshney2016browshing} revealed a phishing vector called \textit{Browshing}, where extensions mimic legitimate sites to steal sensitive information, underscoring the need for targeted phishing detection. Kapravelos et al. \cite{kapravelos2018hulk} examined methods to induce and identify malicious behavior by probing unauthorized data access and page content tampering. Shahriar et al. \cite{shahriar2018effective} proposed a hybrid approach that monitors API calls and user interactions to detect both known and unknown threats. Pantelaios et al. \cite{pantelaios2017youvechanged} developed a method based on analyzing code update deltas to flag potentially harmful modifications, while Moreno et al. \cite{moreno2024vetting} critically assessed the Chrome Web Store's vetting process, revealing that techniques like repackaging and obfuscation can allow malicious extensions to bypass both automated and manual reviews. Despite advancements such as Manifest V3 and sophisticated machine-learning scans, evasion strategies like delayed execution and permission escalation remain effective, highlighting the need for stricter sandboxing, refined permission models, and real-time monitoring.

\section{Advancements and Limitations in Browser Extension Security (2012–2024)}

From 2012 to 2024, major web browsers, including Google Chrome and Mozilla Firefox, have made significant advances in improving extension security to combat persistent threats posed by malicious browser extensions. Initially, both Chrome and Firefox allowed extensive access to browser APIs, making it easier for attackers to exploit vulnerabilities for data theft, phishing, and malware injection \cite{ref8,ref9}. In response to growing threats, Google introduced Manifest V3 (MV3) in 2018, imposing stricter permission requirements, reducing access to sensitive APIs, and replacing the Web Request API with a more restrictive Declarative Net Request (DNR) API. These measures aimed to limit data abuse by malicious extensions while maintaining essential functionality \cite{ref10,ref12}. Additionally, Google improved its Chrome Web Store review process by implementing automated and manual checks to detect malicious behavior before publication \cite{google2024security}. Mozilla Firefox, on the other hand, refined its WebExtensions API, enforcing stricter sandboxing and requiring explicit user consent for sensitive permissions. Unlike Chrome, however, Firefox does not enforce a complete transition to an MV3-like model, keeping the Web Request API accessible, which allows developers greater flexibility but also introduces security risks.

Other browsers, including Microsoft Edge and Safari, have aligned their security models with Chrome and Firefox, adapting their extension frameworks to balance security and developer accessibility. Microsoft Edge, built on Chromium, follows Chrome’s MV3 policies, benefiting from the same security updates. Safari has focused on privacy protections by enforcing stricter permission requests and isolating extensions to limit their access to user data. Despite these advancements, malicious browser extensions continue to evade detection through sophisticated techniques. Extensions often obscure their true intent using obfuscated code or delay activation of malicious behavior until after review \cite{squarex2024}. Cybercriminals frequently update extensions with malicious code or republish previously removed extensions under new names, bypassing detection mechanisms \cite{bleepingcomputer2024chromeextensions}. While Chrome’s MV3 restricts API access, Firefox’s more permissive approach allows for broader functionality, which can be exploited by malicious extensions.

Our research involved creating and testing multiple extensions to evaluate their impact on privacy and security. Minimal permissions—such as \textit{activeTab}, \textit{scripting}, and \textit{storage}—were sufficient for executing harmful actions, making these tools easy to develop even for low-skilled attackers. For example, the \textit{Cookie Stealing} and \textit{Keylogger Extensions} accessed login credentials and cookies, exfiltrating this data to remote servers. The \textit{Screenshot Capture} and \textit{History Tracker Extensions} covertly recorded user activity, offering attackers insights into browsing behavior. Notably, Chrome’s Web Store exhibited stricter security policies, flagging high-risk behaviors like unauthorized cookie access, keystroke logging, and direct DOM manipulation, while Firefox’s Add-ons Store showed greater susceptibility to obfuscation techniques. Extensions designed to modify web content, inject ads, or track user activity were more likely to pass Firefox’s review when disguised as productivity tools.

Despite security enhancements, certain types of extensions, such as those that manipulate browsing history or automate actions like liking content, continue to evade detection. Recent attacks in 2024 further exposed browser vulnerabilities. Extensions like \textit{DataPhisher} and \textit{StealthSpy} bypassed detection with advanced obfuscation, harvesting credentials and manipulating web traffic \cite{bleepingcomputer2024chromeextensions}. While MV3 restricted direct network access in Chrome, attackers leveraged injected scripts and permission abuse to achieve similar results. Firefox’s continued support for the Web Request API increased its exposure to data interception threats. Case studies from 2024 illustrate these risks. \textbf{Cyberhaven} revealed how a supply chain attack compromised developer accounts to distribute extensions that stole Facebook access tokens and bypassed two-factor authentication. \textbf{StealthSpy}, initially marketed as a productivity tool, later functioned as a keylogger, capturing user keystrokes via Chrome’s scripting API. \textbf{AdSkimmer Pro}, disguised as an ad-blocker, injected advertisements and intercepted affiliate revenue, causing financial losses. \textbf{QuickAccess Helper}, promoted as a browsing speed enhancer, redirected users to phishing sites for credential theft.

These incidents highlight the need for stronger security measures across all browsers \cite{cyberhaven2025}, \cite{thehackernews2024}, \cite{arstechnica2020}, \cite{fieldeffect2025}. While Chrome’s MV3 has reduced attack vectors, threat actors continue adapting. Firefox’s lenient API policies present trade-offs between security and flexibility, while Edge and Safari face similar challenges. Continuous improvements in anomaly detection, stricter vetting, and behavioral analysis are essential to mitigating the ongoing risks posed by browser extensions.

\section{Malicious Browser Extensions: Threat Landscape in 2025}
Malicious Browser extensions represent a growing threat to user data privacy and security. This section explores the various types of threats posed by these extensions in the year 2025 and discusses their implications. The study in this section has been done by the authors in their laboratory environment over a period of 6 months. \textcolor{red}{All extensions were created and tested between January and May 2025. While the Chrome and Firefox versions used were released in late 2024, the extensions were submitted to their respective stores in February and March 2025, ensuring that the findings reflect the 2025 threat landscape.}
 The researchers were motivated to build a set of extensions over MV3 for chrome and MV2 for firefox  and using the existing APIs exposed by Chrome and Mozilla till December 2024 to cause a security or privacy issue to the user that installs the extensions. The extensions developed were tested by another researcher and verified to be working fine over their browsers before including them into the set of extensions discussed in this paper. Due to the reason that such extensions can be misused and to not provide an easy-to-build environment of such extensions to script kiddies, only a portion of the codes of the extension is displayed here and no completed code reference has been hosted at any platform. The researchers have studied past work and extensions and tested through experiments whether the threats raised via researchers in the past are still there or are patched and whether there are possibilities of new, more capable malicious extensions that can be developed with the new set of APIs which are available to third-party developers. We have discussed some important malicious browser extensions that we experimented and tested during our study. We classified various extensions that we studied based on the types of threat they pose into the below given 5 classes: 
\begin{itemize}
\item \textbf{Data Stealing Extensions} : Data theft is a primary concern associated with malicious Browser extensions. These extensions can harvest sensitive information such as names, addresses, phone numbers, and email addresses. Furthermore, they are capable of capturing login credentials and stealing financial information such as credit card numbers and bank account details.

\item \textbf{Monitoring and Surveillance Extensions} : Malicious extensions often include monitoring and surveillance capabilities. They can track users' browsing history, record keystrokes through keylogging techniques, and even take screenshots of users' screens without their knowledge or consent. This surveillance can compromise users' privacy and expose sensitive information. Privacy invasion is another significant threat posed by malicious Browser extensions. These extensions may access a user's camera and microphone without authorization, potentially recording audio and video. Additionally, they can track a user's physical location using geolocation APIs, exploiting this information for malicious purposes.

\item \textbf{Content Manipulation Extensions} : Manipulation of web content by malicious extensions is a tactic used to deceive users and achieve nefarious goals. Extensions can inject advertisements into web pages, modify content to mislead users, and employ social engineering techniques to trick users into divulging sensitive information.
\item \textbf{Request Forgery Extensions} : These extensions focus on state-changing actions without user consent. They can manipulate web requests to execute unauthorized actions such as changing user settings, submitting forms, or initiating financial transactions. These activities can lead to unauthorized access, data breaches, and exploitation of users’ online accounts.

\item \textbf{Miscellaneous Extensions} : Malicious Browser extensions are adept at bypassing security mechanisms designed to protect users. They can  avoid detection by security tools, and exploit vulnerabilities in browsers or other software \cite{bleepingcomputer2024chromeextensions}. Furthermore, they may engage in network-based attacks to disrupt users' Internet connections and propagate themselves via social engineering tactics. Additionally, some extensions are specifically designed to manipulate the appearance or functionality of websites, such as altering background colors or injecting hidden elements into the page. While these may seem innocuous at first glance, they can be used for malicious purposes such as redirecting users to fraudulent sites or tricking them into revealing sensitive information. Furthermore, malicious extensions can pose risks through deceptive actions, such as impersonating legitimate tools or services.

\end{itemize}
\section{Experimenting Browser MBEs in 2025}
During our research one of our major contribution is that we have created a set of innovative malicious browser extensions from the past  based on the categorization of malicious actions described in the threat model and realized their execution on the latest and Chrome browser Version 131.0.6778.20 and and Firefox browser Version 123.0. While we discuss the high-level functionality and provide key snapshots of these extensions, we have deliberately withheld the complete code to prevent misuse by malicious actors and script kiddies. To assess whether the created malicious extensions could bypass Chrome Web Store’s and Mozilla Add-On's vetting process, we attempted to submit sample extensions. The extensions that requested excessive permissions or contained obfuscated code were flagged during automated scans. However, those mimicking legitimate functionality with delayed malicious behavior remained undetected, highlighting gaps in security review system. This research provides insights into the evolving threat landscape and highlights how such extensions can compromise privacy and security.

\subsection{Cookie Stealing Extension}
The cookie-stealing extension represents a critical security threat by demonstrating how malicious browser extensions can access and log important access tokens such as cookies without user consent. Attackers can use this method to hijack user sessions, steal authentication tokens, and exfiltrate sensitive information.

\begin{figure}[h]
    \centering
    \includegraphics[width=7.5cm, height=1.6cm]{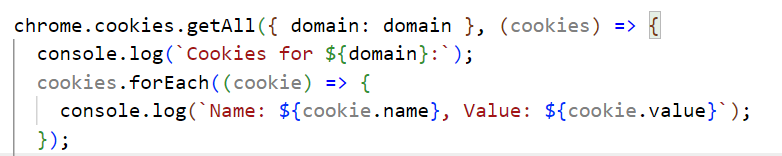}
    \caption{Cookie stealing operation in Chrome and Firefox extensions.}
    \label{fig:cookie}
\end{figure}

The extension operates by listening for specific messages from the browser, particularly those requesting cookie information. Upon receiving such a request, it extracts the domain from the active tab's URL and utilizes the \textit{chrome.cookies. cgetAll} (for Chrome) or \textit{browser.cookies.getAll} (for Firefox) \cite{ref11} API to retrieve all cookies associated with that domain. The extension then logs these cookies, including their names and values, and exfiltrates them to a remote server using a JavaScript Fetch API request. This method allows attackers to gain access to users' authentication tokens, leading to session hijacking and unauthorized account access. The permissions required for this operation include \textit{cookies}, \textit{activeTab}, and \textit{storage} as shown in Figure~\ref{fig:cookie}. 

\subsection{Keylogger Extension}
Keyloggers remain one of the most severe security threats in the digital landscape, as they covertly capture sensitive user inputs, including passwords, personal messages, and financial details. The implementation of a keylogger through a browser extension allows for discreet data collection without the user's awareness.

In this study, we implemented a keylogger using both Chrome and Firefox extensions to analyze its feasibility and impact. The extension works by injecting an event listener into web pages to monitor keypress events. The collected keystrokes are then sent to a background script that processes and logs them. The background script listens for messages from the content script and retrieves recorded keystrokes, which are processed and sent to an external server for storage.

\begin{figure}[h]
    \centering
    \includegraphics[width=9cm, height=3.5cm]{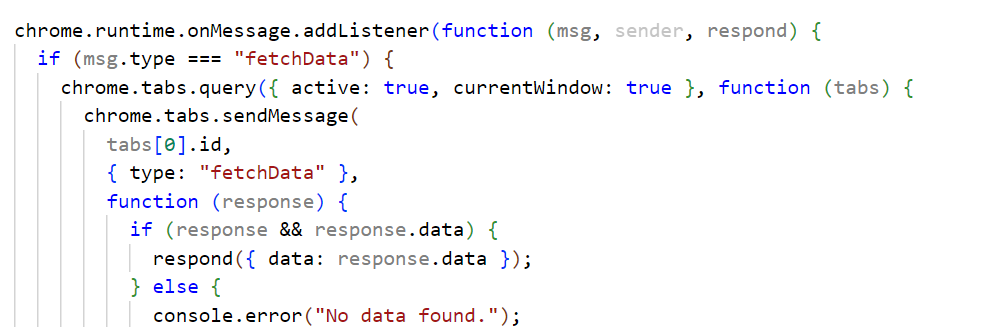}
    \caption{Keylogger operation in Chrome and Firefox extensions.}
    \label{fig:keylogger}
\end{figure}

When a message of type \textit{getKeys} is received, the background script queries the currently active tab using \textit{chrome.tabs.query} (Chrome) or \textit{browser.tabs.query} (Firefox) \cite{ref12}. The recorded keystrokes are then forwarded using \textit{chrome.tabs.send Message} (Chrome) or \textit{browser.tabs.sendMessage} (Firefox) \cite{ref12}, as shown in Figure~\ref{fig:keylogger}. The extension requires \textit{activeTab}, \textit{scripting}, and \textit{storage} permissions.

\subsection{Screenshot Capture by Browser Extension}
Malicious browser extensions can covertly capture screenshots of the active tab, potentially exposing sensitive data such as credentials, financial transactions, and private communications. This capability enables attackers to extract information from restricted web applications and compromise user security.

The extension’s background script initializes upon installation using \textit{chrome.run time.onInstalled.addListener} (Chrome) or \textit{browser.runtime.onInstalled.addListener} (Firefox) \cite{ref15}. The extension monitors tab activity using \textit{chrome.tabs.onUpdated.add Listener} (Chrome) or \textit{browser.tabs.onUpdated.addListener} (Firefox) \cite{ref12}, ensuring it captures screenshots as soon as a user navigates to a new webpage. The captured screenshot is then forwarded to an attacker-controlled server via an HTTP request.

\begin{figure}[h]
    \centering
    \includegraphics[width=10cm, height=3.5cm]{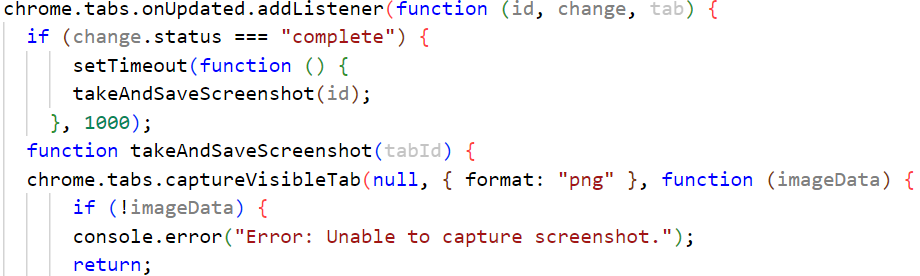}
    \caption{Screenshot capture process in Chrome and Firefox extensions.}
    \label{fig:screenshot}
\end{figure}

The \textit{captureAndDownloadScreenshot} function utilizes \textit{chrome.tabs.captureVisible Tab} (Chrome) or \textit{browser.tabs.captureVisibleTab} (Firefox) \cite{ref12} to take a PNG-format screenshot. The required permissions include \textit{tabs}, \textit{activeTab}, \textit{scripting}, and \textit{storage} as shown in Figure~\ref{fig:screenshot}.

\subsection{History Tracker by Browser Extension}
This extension falls under the category of \textit{privacy invasion} and \textit{data exfiltration attacks}. By leveraging browser APIs, it collects users’ browsing history, which can be used for profiling, targeted phishing, and user behavior analysis.

The extension accesses historical browsing data using \textit{chrome.history.search}, \textit{chrome.history.get Visits} (Chrome) or \textit{browser.history.search}, \textit{browser.history.get Visits} (Firefox) \cite{ref16}. It logs the extracted URLs along with timestamps and user interactions, creating a comprehensive record of the user’s online activity. The data is then transmitted to an external server controlled by the attacker.

\begin{figure}[h]
    \centering
    \includegraphics[width=9cm, height=3.5cm]{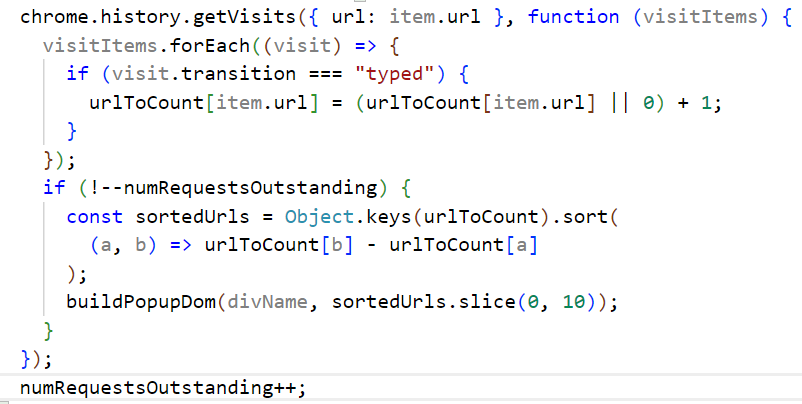}
    \caption{Demonstration of a history-tracking extension in Chrome and Firefox.}
    \label{fig:history}
\end{figure}

The permissions required include \textit{history}, \textit{tabs}, \textit{scripting}, and \textit{storage}. Figure~\ref{fig:history} demonstrates this extension’s activity.

\subsection{Auto-Like YouTube Videos}
The Auto-Like YouTube Videos extension demonstrates how browser extensions can manipulate engagement metrics on social media platforms. This manipulation affects content ranking algorithms and distorts user engagement patterns.

The extension monitors YouTube pages for video elements and automatically triggers a “Like” action when a user hovers over a video thumbnail. It does this using \textit{document.querySelector} to detect the YouTube like button and executes a simulated click event. The extension employs \textit{MutationObserver} \cite{ref18} to ensure the auto-like functionality persists across dynamic page changes. This method can be leveraged by malicious actors to artificially inflate video rankings and influence recommendations.

\begin{figure}[h]
    \centering
    \includegraphics[width=8cm, height=2cm]{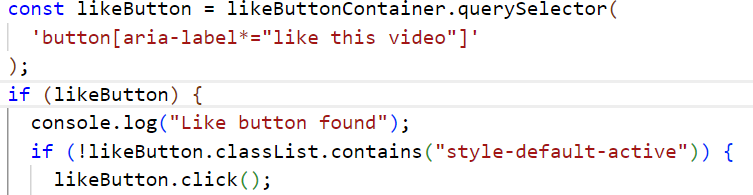}
    \caption{YouTube Auto-Like Extension in Chrome and Firefox.}
    \label{fig:yt}
\end{figure}

The required permissions include \textit{tabs}, \textit{activeTab}, \textit{scripting}, and \textit{storage}. Figure~\ref{fig:yt} illustrates this behavior.

\subsection{Manipulation of Web Content by Browser Extensions}
This extension dynamically modifies web content, posing significant security risks such as phishing, deceptive redirects, and unauthorized content injection. Attackers can use such extensions to alter webpage elements, manipulate hyperlinks, or inject malicious advertisements, leading to data theft or fraud.

\begin{figure}[h]
\centering
\includegraphics[width=7.5cm, height=2.5cm]{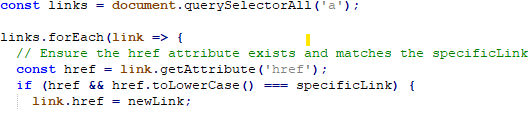}
\caption{JavaScript code snippet for dynamic link modification in Chrome and Firefox extensions.}
\label{fig:code-snippet}
\end{figure}

The extension operates by leveraging the \textit{MutationObserver API} \cite{ref18} to monitor DOM changes. It continuously scans for specific elements, such as anchor tags, and replaces their \textit{href} attributes to redirect users to attacker-controlled domains. This technique allows attackers to conduct phishing attacks by redirecting users to fraudulent login pages or injecting rogue advertisements onto legitimate websites as shown in Figure~\ref{fig:code-snippet}. .

Chrome and Firefox both allow extensions to modify web content, but their implementation of security restrictions varies. Chrome's MV3 model imposes stricter limitations on dynamic script execution, whereas Firefox retains more flexibility, allowing direct script manipulation in certain contexts. This distinction impacts the effectiveness of security policies designed to mitigate such threats.

\subsection{Camera Auto On by Browser Extension}
Unauthorized camera access remains a significant privacy threat posed by malicious browser extensions. This extension demonstrates how an attacker can exploit \textit{navigator.mediaDevices.getUserMedia} to activate a user’s camera without explicit consent.

Upon installation, the extension injects JavaScript into webpages using \textit{chrome. scripting.executeScript} (Chrome) or \textit{browser.scripting.executeScript} (Firefox) \cite{ref12}, ensuring it runs persistently in the background. The extension listens for browser events using \textit{chrome.tabs.onUpdated.addListener} or \textit{browser.tabs.onUpdated.add Listener}, allowing it to activate the camera each time the user loads a webpage. Once triggered, it automatically enables video recording and streams the feed to an external server.

\begin{figure}[h]
\centering
\includegraphics[width=8cm, height=4cm]{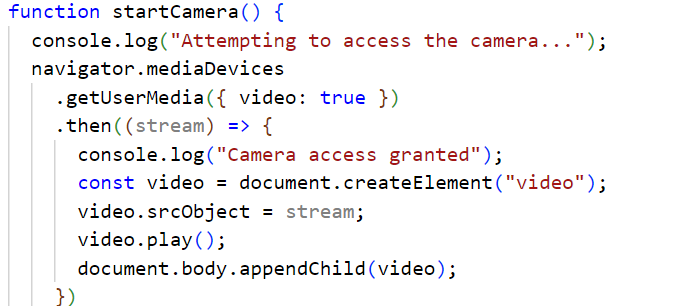}
\caption{Camera Auto On Extension Script Injection in Chrome and Firefox.}
\label{fig:camera_auto_on}
\end{figure}

While Chrome and Firefox enforce user permission requests for media access, extensions with broad permissions can manipulate these settings post-installation, creating a persistent security risk. Figure~\ref{fig:camera_auto_on} illustrates this exploit.

\FloatBarrier

\subsection{Injecting Advertisement through Browser Extensions}
\begin{figure}[htbp]
\centering
\includegraphics[width=8cm, height=3.5cm]{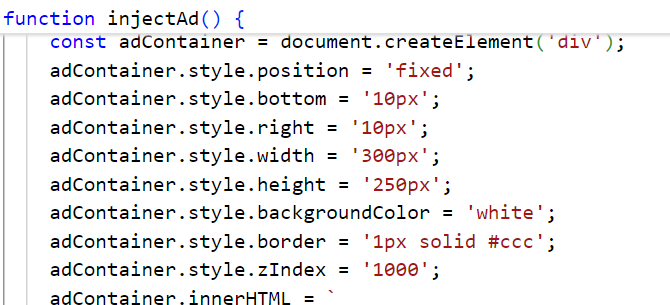}
\caption{JavaScript code snippet for dynamic advertisement injection.}
\label{fig:ad-injection-code}
\end{figure}

One of the browser extensions we analyzed is the ``Inject Advertisement'' extension, which falls under Unwanted Ad Injection attacks. This type of extension disrupts user experience by injecting unauthorized advertisements onto web pages without consent. These ads typically appear as floating banners, pop-ups, or overlays positioned over legitimate content, often redirecting users to external sites, some of which may be fraudulent or malicious. The extension dynamically creates an advertisement container, a \textit{div} element, positioned in the bottom-right corner of the screen. This container contains a header, descriptive text, and a close button, although some malicious versions prevent users from dismissing the ad. The core functionality is implemented via the \textit{injectAd()} function, as shown in Figure~\ref{fig:ad-injection-code}, which uses JavaScript to create and style the advertisement before appending it to the webpage’s DOM. The extension listens for page load events using \textit{window.onload} and employs the \textit{MutationObserver API} to detect dynamic content changes, ensuring the ad persists even when users navigate between pages.

Although the extension only requires the \textit{activeTab} permission, more advanced versions may request \textit{storage} for tracking user interactions, \textit{cookies} for targeted ad injection, and \textit{webRequest} to manipulate network traffic. Such extensions can pose serious security risks, including click fraud, traffic hijacking, and phishing attacks by disguising ads as legitimate notifications or login prompts. Additionally, attackers can replace genuine ads with their own, diverting revenue from website owners. Some variants use drive-by downloads to deliver malware upon interaction with the ad. 
\FloatBarrier

\FloatBarrier
\subsection{Email Inbox Spying by Browser Extensions}This extension highlights the privacy risks posed by unauthorized email monitoring. By exploiting browser APIs, it enables attackers to track unread emails, extract metadata, and manipulate webmail interfaces without the user’s knowledge. The extension operates by monitoring the DOM for elements representing unread emails. It identifies and highlights specific email classes, such as \textit{.zE} in Gmail, using JavaScript. Once identified, it modifies the CSS properties of these elements, making them visibly distinct (e.g., changing the background color to yellow, \textit{\#ffeb3b}). Additionally, the extension logs email metadata and transmits it to an external server, facilitating targeted phishing attacks.

Chrome and Firefox impose different restrictions on DOM manipulation by extensions. While Chrome’s MV3 enforces stricter execution policies, Firefox’s WebExtensions API allows broader modifications, increasing potential security risks.

\begin{figure}[htbp]
\centering
\includegraphics[width=8cm, height=2.5cm]{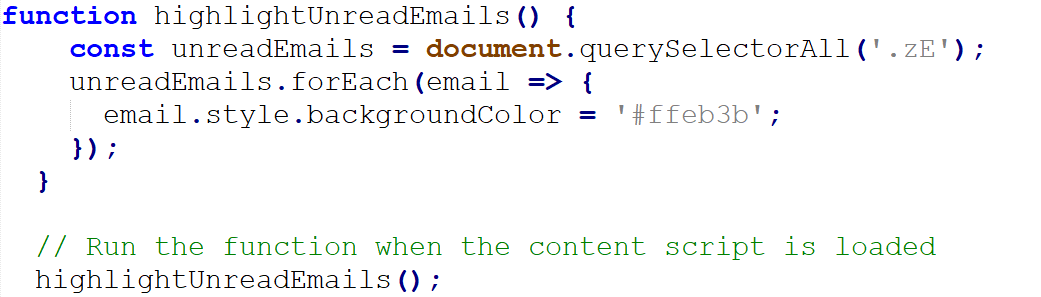}
\caption{JavaScript code snippet for highlighting unread emails in Chrome and Firefox.}
\label{fig:email-spying-code}
\end{figure}

The extension requires minimal permissions—\textit{scripting}, \textit{storage}, and \textit{activeTab} —to perform these operations effectively. Figure~\ref{fig:email-spying-code} illustrates a JavaScript snippet demonstrating this attack.

\begin{table*}[h]
\centering
\caption{Detailed Discussion of Malicious Browser Extensions Created and Their Bypassing Potential}
\label{tab:extensions}
\renewcommand{\arraystretch}{1.4} 
\resizebox{\textwidth}{!}{%
\large 
\begin{tabular}{|p{3cm}|p{6cm}|p{4cm}|p{4cm}|p{4cm}|}
\hline
\textbf{Name} & \textbf{Description} & \textbf{API Exploited} & \textbf{Chrome Web Store} & \textbf{Mozilla Add-on Store} \\ 
\hline

Cookie Stealing Extension & Steals cookies, enabling session hijacking and unauthorized access to user accounts. & \texttt{chrome.cookies. getAll}, Fetch API & Not Possible – Requires \texttt{cookies} permission, detected easily  & Successfully published on Mozilla – Can be obfuscated in a sync manager \\ 
\hline

Keylogger Extension & Captures keystrokes to steal sensitive information such as passwords and personal messages. & \texttt{chrome.tabs.query}, \texttt{chrome.tabs. sendMessage} & Not Possible – Keyboard event tracking flagged  & Successfully published on Mozilla – If embedded in a typing tool \\ 
\hline

Screenshot Capture Extension & Secretly captures screenshots of the user's activity and exfiltrates the data. & \texttt{chrome.tabs. captureVisibleTab}, Fetch API & Not Possible – Screenshots trigger manual review  & Successfully published on Mozilla – Can justify as a productivity tool \\ 
\hline

History Tracker Extension & Monitors and logs browsing history, enabling user profiling. & \texttt{chrome.history. search}, \texttt{chrome.history .getVisits} & Not Possible – Browsing history requires permissions  & Successfully published on Mozilla – If marketed as “Browsing Insights” \\ 
\hline

Auto-Like YouTube Videos Extension & Automatically likes videos, skewing engagement metrics. & MutationObserver, \texttt{document.query Selector} & Successfully published on Chrome – If embedded in focus tools\cite{youtube_focus_timer}  & Successfully published on Mozilla – Hidden in productivity tools\cite{firefox_focus_timer} \\ 
\hline

Manipulation of Web Content Extension & Alters page content to facilitate phishing or redirection. & MutationObserver, \texttt{chrome.scripting. executeScript} & Not Possible – DOM modifications trigger review  & Successfully published on Mozilla – Disguised as a UI enhancement \\ 
\hline

Camera Auto-On Extension & Activates the user's camera without consent for surveillance. & \texttt{navigator. mediaDevices. getUserMedia}, \texttt{chrome.scripting. executeScript} & Not Possible – Direct camera activation blocked  & Successfully published on Mozilla – Possible with delayed triggers \\ 
\hline

Injecting Advertisement Extension & Injects unwanted ads that may lead to malware or phishing. & \texttt{document. createElement}, \texttt{document.body. appendChild} & Possible – Can be hidden in UI tweaks  & Successfully published on Mozilla – As a customization tool\cite{firefox_easy_todo} \\ 
\hline

Email Inbox Spying Extension & Monitors unread email indicators and exfiltrates data. & \texttt{document. querySelector}, \texttt{chrome.scripting. executeScript} & Not Possible – Gmail tracking gets flagged  & Successfully published on Mozilla – Blended into “Email Organizer” \\ 
\hline

\end{tabular}%
}
\end{table*}

\subsection{Analysis and Observations}
Our research involved creating and testing multiple malicious browser extensions to evaluate their impact on privacy and security. A key finding is that minimal permissions—such as \textit{activeTab}, \textit{scripting}, and \textit{storage}—are sufficient for executing harmful actions, making these tools easy to develop even for low-skilled attackers. For example, the \textit{Cookie Stealing} and \textit{Keylogger Extensions} accessed sensitive data like login credentials and cookies, then exfiltrated this information to remote servers. Similarly, the \textit{Screenshot Capture} and \textit{History Tracker Extensions} covertly recorded user activity and captured screenshots, providing attackers with detailed insights into browsing behavior. When evaluating the security measures of Chrome and Firefox extension stores, we observed notable differences in their ability to detect and block malicious extensions. Chrome’s Web Store exhibited stricter security policies, flagging high-risk behaviors such as unauthorized cookie access, keystroke logging, and direct DOM manipulation. Extensions attempting to modify browsing history or inject advertisements were often detected and removed during the review process. On the other hand, Firefox’s Add-ons Store demonstrated greater susceptibility to bypass techniques, particularly when malicious behavior was obfuscated within extensions that appeared to provide legitimate functionality. For instance, extensions that manipulated web content, injected ads, or tracked user activity were more likely to pass Firefox’s review when disguised as productivity tools or interface enhancements. Additionally, we successfully created a \textit{To-Do List Extension} on Mozilla that contained obfuscated malicious code to inject advertisements, which successfully bypassed Mozilla's review process and remains active on the Add-ons Store \cite{firefox_easy_todo}. Similarly, we developed a \textit{25-Minute Timer Extension} on Mozilla that included hidden functionality for automatically liking YouTube videos, demonstrating that extensions with benign primary functions can effectively disguise malicious intent and evade detection \cite{firefox_focus_timer}.
On Chrome, we created an extension that displayed a pop-up showing YouTube access time while secretly implementing an \textit{auto-liking mechanism for YouTube videos}. This extension was successfully published on the Chrome Web Store and remains active, highlighting weaknesses in Chrome’s automated review process for behavioral detection \cite{youtube_focus_timer}. \textcolor{red}{Note: All extensions and code snippets described in this section are original implementations developed by the authors solely for research purposes.}

Despite Chrome’s stronger enforcement mechanisms, certain types of extensions, such as those designed for auto-liking content and modifying browsing history, could still evade detection by embedding their malicious logic within seemingly harmless scripts. Similarly, Firefox's review process was found to be more lenient toward extensions with broad permissions, allowing them to perform unauthorized data collection and behavioral tracking under the guise of enhancing user experience. The \textit{Camera Auto-On Extension} further demonstrated the risks posed by persistent permissions, as it was capable of secretly activating a user's camera on both browsers, albeit with a higher likelihood of detection in Chrome due to its stricter permission review system.

Table~\ref{tab:extensions} provides a comparative analysis of these extensions, detailing their capabilities, exploited APIs, required permissions, and the likelihood of bypassing security measures in Chrome and Firefox. Although both browsers employ security mechanisms, the deceptive tactics used by these extensions highlight significant gaps in the review process, particularly in Firefox’s ability to detect obfuscated threats. These findings emphasize the need for more rigorous permission validation, improved static and dynamic analysis tools, and enhanced monitoring mechanisms to prevent unauthorized access to user data and mitigate the risks posed by malicious extensions. \textcolor{red}{While we have not yet submitted formal disclosures to Mozilla and Chrome regarding our test extensions (e.g., auto-like or ad-injecting behaviors), we recognize the importance of responsible disclosure. We plan to report these results to the appropriate browser security teams prior to final publication, in line with ethical research standards.}

\section{Conclusions}
Browser extensions, while offering enhanced functionality, pose a significant threat to user privacy and security due to their vulnerabilities. This study has demonstrated how malicious browser extensions can exploit minimal permissions to execute attacks such as data theft, surveillance, and unauthorized content manipulation. A key finding is the disparity between Chrome and Firefox in detecting and mitigating these threats. While Chrome's Web Store enforces stricter security measures, blocking many high-risk behaviors such as unauthorized cookie access and keystroke logging, Firefox’s Add-ons Store remains more susceptible to bypassing techniques, particularly when malicious behavior is obfuscated within seemingly benign extensions.

Alarmingly, despite extensive research, the solutions proposed by the academic and cybersecurity community are rarely implemented in real-world scenarios. The ability of attackers to disguise malicious intent within productivity or customization tools highlights critical weaknesses in the current extension vetting process. Moreover, creating and distributing malicious extensions remains alarmingly simple, and current security measures fail to address the risks posed by post-publication modifications. Once an extension is approved and published, there is little oversight ensuring that subsequent updates do not introduce malicious functionalities. This gap in continuous security monitoring exposes users to persistent cyber threats.

Addressing these challenges requires a multi-faceted approach. Browser vendors must enforce stricter policies, particularly in Firefox’s review process, by integrating enhanced static and dynamic analysis techniques capable of detecting obfuscation and hidden payloads. Additionally, real-time monitoring mechanisms should be implemented to detect behavioral anomalies even after an extension has been approved. Ensuring that updates to extensions undergo rigorous security checks, rather than relying solely on pre-approval evaluations, is crucial for mitigating post-publication risks. Simultaneously, users must be educated on the dangers posed by browser extensions and encouraged to grant permissions judiciously.


Future research should focus on improving real-time behavioral analysis of browser extensions, refining permission models, and establishing industry-wide vetting standards to mitigate these persistent threats. In particular, developing sophisticated anomaly detection techniques and fostering collaboration between browser vendors, security researchers, and policymakers can enhance the effectiveness of existing defenses. Standardizing permission transparency and incorporating automated rollback mechanisms for post-publication updates will be critical steps in strengthening browser extension security. \textcolor{red}{While our experiments focused on Chrome and Firefox on Windows/Linux environments, future work will aim to replicate these findings on macOS and alternative browsers such as Safari and Edge to better generalize the threat landscape across platforms.} Ultimately, bridging the gap between theoretical security measures and their real-world implementation is essential. Without proactive efforts from the industry and researchers, malicious extensions will continue to exploit vulnerabilities, posing a persistent threat to the digital ecosystem. \textcolor{red}{Based on our experimental analysis, we propose the following concrete defenses: (1) sandboxed testing environments for pre-publication review, where extensions are executed in isolated sessions to detect hidden behavior; (2) integration of machine learning models that analyze extension behavior over time to catch delayed-onset malicious activities; (3) implementation of permission change alerts to notify users when an extension updates and requests new capabilities; and (4) post-publication telemetry-based monitoring to identify anomalous network or DOM behaviors. These measures, grounded in our findings, can greatly strengthen the browser security ecosystem.}

\end{document}